\documentclass[12pt]{article}
\usepackage{amsfonts}
\usepackage{amssymb}
\usepackage{color}
\usepackage{graphicx}
 \usepackage{subfigure}
 \usepackage{graphics,epstopdf}
\usepackage{amsthm}
\usepackage{amsmath}

\newcommand{\Sc}{{\cal S}}

\newcommand{\mc}{\mathcal}

\newcommand{\be}{\begin{equation}}
\newcommand{\en}{\end{equation}}
\newcommand{\bea}{\begin{eqnarray}}
\newcommand{\ena}{\end{eqnarray}}
\newcommand{\beano}{\begin{eqnarray*}}
\newcommand{\enano}{\end{eqnarray*}}

\newcommand{\A}{{\cal A}}
\newcommand{\B}{{\cal B}}
\newcommand{\M}{{\cal M}}
\newcommand{\C}{{\cal C}}

\renewcommand{\l}{\langle}
\renewcommand{\r}{\rangle}
\newcommand{\pin}[2]{\l#1 , #2\r}

\newcommand{\Lc}{{\cal L}}
\newcommand{\1}{1 \!\! 1}

\newcommand{\Pc}{{\mc P}}

\newcommand{\Hil}{\mc H}

\catcode `\@=11 \@addtoreset{equation}{section}

\catcode `\@=12
\begin{document}

\vspace*{-2cm}
\vspace{1cm}
\noindent This is the pre-peer reviewed version of the article published in final form at\\ https://doi.org/10.1111/sapm.12566. \\
%

\begin{center}
{\Large \textbf{Dynamics for a quantum parliament}} \vspace{2cm%
}\\[0pt]

{\large F. Bagarello}
\vspace{3mm}\\[0pt]
Dipartimento di Ingegneria,\\[0pt]
Universit\`a di Palermo,\\ I-90128  Palermo, Italy\\
and I.N.F.N., Sezione di Catania

\vspace{7mm}

{\large F. Gargano}
\vspace{3mm}\\[0pt]
Dipartimento di Ingegneria,\\[0pt]
Universit\`{a} di Palermo, I - 90128 Palermo

\end{center}

\vspace*{2cm}

\begin{abstract}
\noindent In this paper we propose a dynamical approach based on the Gorini-Kossakowski-Sudarshan-Lindblad equation for a problem of decision making. More specifically, we consider what was recently called {\em a quantum parliament}, asked to approve or not a certain law, and we propose a model of the connections between the various members of the parliament, proposing in particular some special form of the interactions giving rise to a {\em collaborative} or {\em non collaborative} behaviour.
\end{abstract}

\vfill

\newpage


\section{Introduction}

As it is well established since many centuries, the role of mathematics is relevant not only  in pure, but also in applied sciences. Classical mechanics is a prototypical example of this claim: we know very well that mathematics, and mathematical modelling in particular, is essential if we want to describe the motion of macroscopic systems, like pendulums, bullets, spinning tops and so on. And mathematics is essential also in quantum mechanics, which is all based on some basic facts in functional analysis, Hilbert spaces, and operator theory.

Quantum mechanics is usually associated to the microscopic world, like atoms and molecules, for instance. However, it is now well diffused the feeling that quantum tools, and quantum ideas, can be relevant also for describing some macroscopic systems. Nowadays there are hundreds of papers which explore these connections, and several monographs: We only cite the latter here, where several other references can also be found: \cite{baa}-\cite{bagbook2}.

In relation with our paper, quantum techniques have been applied to political systems with different problems in mind: for instance, in \cite{bag2015SIAP,baggargPhysA} the problem of alliances was considered, {whilst in \cite{pol2016} competition and cooperation were analyzed in a "coalition" equilibrium model}. In other papers, \cite{pol1}-\cite{pol5}, the role of quantum protocols in a voting process has been discussed, while in \cite{rosa1,rosa2} the attitude of people elected in one party to move to a different one has been analyzed.

Recently, \cite{andro}, the interest was more focused on the behaviour of two or three groups of legislators, members of what the authors have nicely  called a {\em quantum parliament}. They need to vote for a given law, which should be accepted or refused, and, while taking their decision, they follow their leader's suggestions. But not completely. In other words, they have a sort of {\em free will}, which each legislator experience while producing its own decision. 

In this paper we consider the same problem as in \cite{andro}, but we adopt a different, dynamical, strategy based on the use of the Lindblad operators. We suppose that  each legislator $\Lc_j$, $j=1,2,\ldots,N$, where $N$ is the total number of the members of the parliament $\pi$, is described by a quantum state, and we give a time dependence to the state by means of a master equation simulating the interaction of $\Lc_j$ with the other members of $\pi$, using both a suitable Hamiltonian and a Lindblad operator to model the possible influences of a leader's party and other effects. The result of these interactions is reflected in the time evolution of the state, and consequently in the final decision of $\Lc_j$. As in \cite{andro}, we consider in our treatment a sort of {\em free will} in $\pi$. This introduces some alea in the final decision of each $\Lc_j$, which, for this reason, is  not  obvious a priori, so that the acceptance of the law is not granted. Our interest is mainly in the derivation of the outcome of the vote, using a minimal set of working assumptions.

The paper is organized as follows: in Section \ref{sect2} we introduce the problem and discuss some of the essential aspects of the framework used all along the paper. In Section \ref{sect3} we discuss the dynamics of our quantum parliament, considering different simplifying situations, from a single party with its leader, to the case of three parties with different peculiarities. The case of more leaders of the same party is also considered. Section \ref{sectconclusions} contains our conclusions and some plans for the future.

\section{Stating the problem}\label{sect2}

Let $\pi$ be the whole parliament. Following \cite{andro} we consider three groups of legislators (or agents) in $\pi$, $\Pc_\A$, $\Pc_\B$ and $\Pc_\M$. These are Alice's party, made of $n_\A$ agents,  Bob's party, made of $n_\B$ legislators, and the {\em mixed group}, consisting in $n_\M$ legislators. We have $n_\A+n_\B+n_\M=N$, the total number of the members of $\pi$. We call $p_\C(j)$ the $j$-th member of the group $\C$, where $\C=\A,\B,\M$ and $j=1,2,\ldots,n_\C$. The difference between the parties is as follows: suppose $\pi$ have to decide whether to accept or refuse a law, $\Lambda$. Alice wants $\Lambda$ to be accepted: she says {\em yes} to $\Lambda$.  On the other side Bob, and its group $\Pc_\B$ can in principle vote "yes" or "no" depending on the interaction with $\Pc_{\A}$. However, we will assume later that Bob is against $\Lambda$. The case with no interaction is almost trivial, since in this case $\Pc_\B$ is essentially a copy of $\Pc_\A$, and we could simply {\em double} the analysis given in Section \ref{sectIII1} below where the influence of the leader, Alice or whoever, on its party has been analyzed in detail. More interesting is the case in which $\Pc_{\A}$ and $\Pc_\B$ interact. And even more interesting is the introduction of a third party, $\Pc_\M$, which we suppose to have no leader and no a priori position to follow.  

Since the only possible choices for $\Lambda$ are "yes" and "no", it is natural to imagine that each $p_\C(j)$ is described by a linear combination of two orthogonal vectors, one representing the choice "yes", the vector $e_0^\C(j)$, and the vector $e_1^\C(j)$, corresponding to the choice "no". These two vectors form an orthonormal (o.n.) basis in the Hilbert space $\Hil_j^\C=\mathbb{C}^2$, endowed with its standard scalar product $\pin{.}{.}_j$. The general vector of $p_\C(j)$ can be written as
\be
\psi_j^\C=\alpha_j^\C e_0^\C(j)+\beta_j^\C e_1^\C(j),
\label{21}\en
with $|\alpha_j^\C|^2+|\beta_j^\C|^2=1$. Here, as before,  $j=1,2,\ldots,n_\C$ and $\C=\A,\B,\M$. $|\alpha_j^\C|^2$ and $|\beta_j^\C|^2$ can be seen respectively as the probability of $p_\C(j)$
to vote "yes" or "no".
We can naturally associate a composite Hilbert space to each of the parties in $\pi$, made of copies of $\mathbb{C}^2$. In particular we put
$$
\Hil_\A=\otimes_{j=1}^{n_\A}\Hil_j^\A, \qquad \Hil_\B=\otimes_{j=1}^{n_\B}\Hil_j^\B, \qquad \Hil_\M=\otimes_{j=1}^{n_\M}\Hil_j^\M,
$$
and
\be
\Hil=\Hil_\A\otimes\Hil_\B\otimes\Hil_\M.
\label{hisbertspace}\en
An o.n. basis of $\Hil$ is clearly consisting of tensor products of states $e_0^\C(j)$, and $e_1^\C(j)$ for various $j$ and different $\C$, that is of the vectors of the set
\bea
\mathcal{E}=\{e_0^\A(1),e_1^\A(1),\ldots,e_0^\A(n_\A),e_1^\A(n_\A),e_0^\B(j),e_1^\B(1),\dots,e_0^\M(n_\M),e_1^\M(n_\M)\}.\ena

 For instance, a vector describing a situation in which all the legislators of $\Pc_\A$ and $\Pc_\M$ vote "yes", while all those of $\Pc_\B$ vote "no" is
$$
(e_0^\A(1)\otimes\cdots\otimes e_0^\A(n_\A))\otimes(e_1^\B(1)\otimes\cdots \otimes e_1^\B(n_\B))\otimes(e_0^\M(1)\otimes\cdots \otimes e_0^\M(n_\M)).
$$
Of course the dimensionality of $\Hil$ increases with $N$. In fact we have $\dim(\Hil)=2^N$. An operator $X$ acting on, say, $\Hil_1^\A$, is identified with the tensor product $X\otimes \mathcal{I}_2\otimes\cdots\mathcal{I}_2$, i.e. the tensor product of $X$ with $N-1$ copies of the identity operator $\mathcal{I}_2$, acting on all the other Hilbert spaces.
However,  see Sections II and III, in our applications we shall consider some simplifications that reduce significantly the dimensionality of the Hilbert space. In particular, we will suppose that all the members of a specific party are {\em indistinguishable}, i.e., they all feel the same interactions and are described by the same parameters, with the same values, so that it is reasonable to restrict to $n_\A=n_\B=n_\M=1$, and the various $\Hil_\A,\Hil_\B,\Hil_\M$ have dimension 2, while $\Hil$ has dimension 8. This is why, in the rest of the paper, we will often focus on a single legislator of $\pi$, $\Lc$, or, at most, on a single representative member for each party.  

We consider now the vector
\be
\psi=\alpha\, e_0+\beta\, e_1,
\label{22}\en
with $|\alpha|^2+|\beta|^2=1$. This $\psi$ is, of course, analogous to the one in (\ref{21}) adopting a simplifying notation which is sufficient now since we are only focusing on $\Lc$. We should also mention that, in the following, sometimes we will use $|e_j\rangle$ rather than simply $e_j$, $j=0,1$. This is useful when, see for instance  $\rho_\psi$ below, rather than to the vectors, we deal with density matrices.
A possible parametrization of the vector $\psi$, adopted in particular in \cite{andro}, is the following: $\alpha=\cos(\theta/2)$ and $\beta=\sin(\theta/2)\,e^{i\varphi}$, with $\theta\in[0,\pi]$, and $\varphi\in[0,2\pi[$. 

We can now define a density matrix $\rho_\psi$ as an operator acting on $\Hil=\mathbb{C}^2$ as follows: $\rho_\psi f=\pin{\psi}{f}\psi$, $\forall f\in\Hil$. In the bra-ket language, $\rho_\psi$ is often written as $\rho_\psi=|\psi\rangle\langle\psi|$. More explicitly, $\rho_\psi$ is the two-by-two matrix
\bea
\rho_\psi=\left(
\begin{array}{cc}
	|\alpha|^2  & \alpha\,\overline\beta\\
	\overline\alpha\,\beta  & |\beta|^2\\
\end{array}
\right),\label{23}
\ena
which is manifestly self-adjoint and with unit trace. In particular, if $\Lc$ is in a "yes" or in a "no" {\em mood}, $e_0$ or $e_1$, the related density matrices are
\bea
\rho_0=\left(
\begin{array}{cc}
	1  & 0\\
	0  & 0\\
\end{array}
\right), \qquad \rho_1=\left(
\begin{array}{cc}
	0  & 0\\
	0  & 1\\
\end{array}
\right),\label{density}
\ena
with obvious notation.
In \cite{andro} the main idea was to {\em measure} the distance of the density matrix of the various legislators with those corresponding to Alice ($\rho_0$) and Bob ($\rho_1$). This is a way to check which the final decision of each legislator is. Moreover, as already observed, a {\em free will parameter} was introduced in \cite{andro} for $\Pc_\A$ and $\Pc_\B$. For instance, let $r_\A$ be this parameter for $\Pc_\A$. Then, it is not required that $\rho_\psi=\rho_0$  to conclude that $\Lc$ is going to vote "yes". It is sufficient that the difference between $\rho_\psi$ and $\rho_0$ is less than $r_\A$ or, more explicitly, that $d(\rho_\psi,\rho_0)\leq r_\A$. Here $d(.,.)$ is a {\em distance} between density matrices. Like any distance, $d(.,.)$ must satisfy some constraints: it must be symmetric, $d(\rho_\psi,\rho_\varphi)=d(\rho_\varphi,\rho_\psi)$, non negative, $d(\rho_\psi,\rho_\varphi)\geq0$, with in particular $d(\rho_\psi,\rho_\varphi)=0$ if and only if $\rho_\psi=\rho_\varphi$, and it must satisfy the triangular inequality: $d(\rho_\psi,\rho_\varphi)\leq d(\rho_\psi,\rho_\eta)+d(\rho_\eta,\rho_\varphi)$, for all density matrices $\rho_\psi$, $\rho_\eta$ and $\rho_\varphi$. In \cite{andro} the explicit expression for the distance was the following:
$$
d(\rho_\psi,\rho_\varphi)=\frac{1}{2}\,\text{Tr}|\rho_\psi-\rho_\varphi|,
$$
where $\text{Tr}|A|$ is the trace and $|A|=\sqrt{A^\dagger A}$, $A\in \M_2$, the set of all two-by-two matrices. Here we just want to briefly comment that, in our opinion, this choice of distance does not always perform well, at least if we still want to give a full meaning to the previous parametric representation for $\alpha$ and $\beta$. In fact, using this parametrization, we get $d(\rho_\psi,\rho_0)=\sin(\theta/2)$ and $d(\rho_\psi,\rho_1)=\cos(\theta/2)$, which are both independent of $\varphi$. This is reasonable when we use density matrices, but not so much when we adopt vectors, which is also a natural choice, used several times in the literature in similar situations. In this latter case, the distance $d_n(\psi,\phi)=\|\psi-\phi\|$, $\forall\psi,\phi\in\Hil$, could be a good alternative, having no drawback of the kind shown for $d(.,.)$.


Our approach is  based on the derivation of suitable mean values of some observables describing the final decision of the legislator $\mathcal{L}$. In particular, using the  density matrices defined in \eqref{density}, their mean values are obtained through
	\bea
	\langle\rho_{0}\rangle_{\psi}=\textrm{Tr}\left[\rho_{\psi}\rho_0\right],\qquad	\langle\rho_{1}\rangle_{\psi}=\textrm{Tr}\left[\rho_{\psi}\rho_1\right].\label{meanrhob}
	\ena
	 As we shall see, the above expressions can be straightforwardly extended to take into account the presence of multiple members belonging to different parties (see for instances \eqref{meanrho} below).  The meaning of \eqref{meanrhob} is quite evident: $\langle\rho_{0}\rangle_{\psi}$ is the mean value of the operator $|e_0\rangle\langle e_0|$ which represent the pure state corresponding to "$\Lc$ votes yes", whereas 
	 $\langle\rho_{1}\rangle_{\psi}$ is the mean value of $|e_1\rangle\langle e_1|$, the pure state representing, this time, "$\Lc$ votes no". Both these mean values are computed on the density matrix $\rho_\psi$ describing $\Lc$. Using the properties $\text{Tr}[\rho_\psi]=1$ and $\rho_0+\rho_1=\1$, one can easily obtain $\langle\rho_{0}\rangle_{\psi}=1-\langle\rho_{1}\rangle_{\psi}$.  Hence, the time evolution of $\langle\rho_{0}\rangle_{\psi}$ and $\langle\rho_{1}\rangle_{\psi}$ can be phenomenologically interpreted as a measure of the legislator's decision to vote "yes" or "no".

\section{The dynamics of the system}\label{sect3}

Our main effort consists in proposing a plausible dynamics for the generic member of the group $\cal C$. This means that we are supposing that the member's decision  can change in time due, for instance, to the parties' influence or to the presence of one or more leaders, not necessarily belonging to different parties.
The original vector (\ref{22}) $\psi$ becomes now time-dependent, $\psi(t)$, and this new vector still  belongs to the same Hilbert space $\Hil$ in (\ref{hisbertspace}): again we omit the label of the party since the mechanisms we shall describe are essentially the same for each legislator, and we are focusing on just one of them, $\Lc$.
 
It is well known that the dynamics of the wave function of a closed quantum system is governed by the  Schr\"{o}dinger equation:
\bea
i\,\frac{d}{d t}\psi(t)= H\psi(t),\label{sh}
\ena
where  $H=H^\dag$ is the self-adjoint Hamiltonian operator containing all mechanisms acting in the closed system. Given the initial condition $\psi(0)$ the evolution $\psi(t)$ is completely determined. Formally we have $\psi(t)=e^{-iHt}\psi(0)$.

However our system can be seen as open, where the {\em small} closed subsystem is made by  the various members of the parties, while their reservoirs are nothing but the parties themselves. This is because the parties can influence the single legislator, while the opposite is quite less plausible. Stated differently, Alice (resp. Bob) influences what each single $p_\A(j)$ (resp. $p_\B(j)$) decides, but $p_\A(j)$'s (resp. $p_\B(j)$'s) opinion is not relevant for Alice (resp. Bob). This kind of one-directional flow of influence suggests that the dynamics in (\ref{sh}) is not the most appropriate, then. The reason is twofold: first of all, self-adjointess of $H$ does not allow, by itself, to avoid identical strength of interactions between, say, Alice and $p_\A(j)$: if $H=H^\dagger$, then if Alice {\em communicates} with $p_\A(j)$ with a given strength, then  $p_\A(j)$ {\em communicates back}  with Alice with the same strength. Secondly, it has been proved in recent years, see \cite{bagbook2} for instance, that $H=H^\dagger$ is only compatible, if $dim(\Hil)<\infty$, as in our case, with periodic or quasi-periodic dynamics. But such an oscillating dynamics, of course, is not really what one expects in a decision-making process, where one imagines to find some limiting value, corresponding to the {\em final decision}. Hence,
 we need to include some non-Hermitian effect, that we mimic here trough the Gorini-Kossakowski-Sudarshan-Lindblad (GKSL) equation, see for instance \cite{Lin,Bre} for open quantum systems,  and \cite{Nava2022,khren1,Asan2013} for applications outside the quantum realm.

 \vspace{2mm}
 
 {\bf Remark:--} The GKSL equation below is not the only possibility to analyse the time evolution of a given open quantum system. Other possibilities are also well known, as a purely Hamiltonian approach in which the Hamiltonian includes also terms of the reservoir, \cite{bagbook,bagbook2,Bre}, or using the so-called $(H,\rho)$-induced dynamics, \cite{BDGO}.
 
 \vspace{2mm}

In particular, considering the  density operator $\rho(t)=|\psi(t)\rangle\langle\psi(t)|$ whose matrix representation is \eqref{23}, but with time dependent parameters,  the resulting differential equation for the evolution of $\rho$ is the well known GKSL equation:

\bea
\frac{d}{d t}\rho(t)=-i [H, \rho(t)]+\sum_{j=1}^{N}\left(L_{j} \rho(t) L_{j}^{\dagger}-\frac{1}{2}\left\{L_{j}^{\dagger} L_{j}, \rho(t)\right\}\right),\label{eq:lineq}
\ena
where the various  $L_{j}$ are  the Lindblad operators,  generally taken traceless and connected to the influence of the larger system (the parties) on the small system (the various members of $\pi$), {and where $\{X,Y\}=XY+YX$ is the anti-commutator between two generic operators $X$ and $Y$.}
{In writing this equation we have assumed that the dynamics of the system is Markovian, and the Lindblad operators (which describe the way the Leader influences the various members of his party) are independent of the current state of the system. This can be simply understood under the assumption that the leader is not influenced in any way by the reaction of the members. This requirement is a natural way of interpreting an environment which is negligibly perturbed by a {\em small} system. Of course this is a strong hypothesis that however is plausible in a dynamics ruled by some leadership.}
We notice that when the interaction with the larger system is not included, we recover the Von Neumann equation:
\beano
\frac{d}{d t} \rho(t)=-i [H, \rho(t)],
\enano
which could be easily deduced from \eqref{sh}, since $\rho(t)=|\psi(t)\rangle\langle\psi(t)|$.

For the interpretation of our model it is useful to remind that the interaction between the small system with the environment produces a mixture of states from a generic pure state, \cite{Manz}.
In fact, {adopting a standard perturbative approach for small times} and neglecting for a moment the action of the Hamiltonian\footnote{We are assuming that its effect is negligible with respect to the Lindbladian part, so that the Markovian process is highlighted.} $H$ in (\ref{eq:lineq}),  the evolved density operator of a pure state $\rho=|\psi\rangle\langle\psi|=\rho(0)$ in a small time step $dt$ can be rewritten, to the leading order in $dt$, as

\bea
\rho(dt)\approx \rho-\frac{1}{2}dt\sum_{j=1}^N\left(L^\dagger_jL_j\rho+\rho L^\dagger_jL_j\right)+dt\sum_{j=1}^N L_j\rho L^\dagger_j\approx
\mathcal{A}\, \rho\,\mathcal{A}^\dagger+\sum_{j=1}^N\mathcal{B}_j\, \rho\,\mathcal{B}_j^\dagger, \label{linjumps}
\ena
where 
\bea
\mathcal{A}=\1-\frac{dt}{2}\sum_{j=1}^{N}L_{j}^{\dagger}L_{j},\qquad
 \mathcal{B}_j=\sqrt{dt}\,L_j,\,j=1,\ldots,N.
\ena
In other words, the evolved state is a mixture of the pure states defined by
$\mathcal{A}|\psi\rangle$ and by the various $\mathcal{B}_j|\psi\rangle$.
In particular we have
\beano
\mathcal{A}\,\rho\,\mathcal{A}^\dagger=\mathcal{A}\, |\psi\rangle\langle\psi|\,\mathcal{A}^\dagger=p_{\mathcal{A}}
\mathcal{\tilde A}\, |\psi\rangle\langle\psi|\mathcal{\tilde A}^\dagger,\label{Aj}
\enano
with $\mathcal{\tilde A}=\frac{1}{\|\mathcal{ A|\psi\rangle\|}}\mathcal{A}$. Here $p_{\mathcal{A}}={\|\mathcal{A}\, |\psi\rangle\|}^2\simeq\left(1-dt\sum_{j=1}^N\|L_j|\psi\rangle\|^2\right)$ can be seen as the probability that 
the vector $\psi$ evolves in $\mathcal{\tilde A}\, |\psi\rangle$. 
This vector for $dt\rightarrow0$ tends to the initial vector $|\psi\rangle$, and follows the so called  \textit{continuous drift-type evolution}, \cite{Manz}.

With similar computations we have
\bea
\mathcal{B}_j\, \rho\,\mathcal{B}_j^\dagger=\mathcal{B}_j\, |\psi\rangle\langle\psi|\,\mathcal{B}_j^\dagger=p_{\mathcal{B}_j}
\mathcal{\tilde B}_j\, |\psi\rangle\langle\psi|\mathcal{\tilde B}_j^\dagger,\,\forall j=1,\ldots,N,\label{Bj}
\ena
where   $\mathcal{\tilde B}_j=\frac{1}{\|\mathcal{ B}_j|\psi\rangle\|}\mathcal{B}_j$ and $p_{\mathcal{B}_j}={\|\mathcal{B}_j\, |\psi\rangle\|}^2=dt\|L_j|\psi\rangle\|^2$ is the probability that 
the vector $\psi$ evolves in $\mathcal{\tilde B}_j\, |\psi\rangle$\footnote{Notice that due to \eqref{linjumps}, here we simply have $\mathcal{\tilde B}_j=\frac{1}{\|\mathcal{L}_j|\psi\rangle\|}\mathcal{L}_j$}.
The process of evolving in such a state is called \textit{evolutionary jumps} as for $dt\rightarrow0$ the vector $\mathcal{\tilde B}_j\, |\psi\rangle$ does not tend to the original $|\psi\rangle$. This is the key process that produces mixed states as a consequence of the interaction of the system with the large environment. As it is well known, this process can be detected by looking at the so-called purity, $\mathcal{P}=\text{Tr}(\rho^2)$, and at the Von Neumann entropy 
\bea \mathcal{S}=-\text{Tr}(\rho \log\rho)\label{entro},\ena that for a mixed state satisfy the inequalities $\mathcal{P}<1,\, \mathcal{S}>0$,  while  $\mathcal{P}=1$ and $\mathcal{S}=0$ for pure states.

\subsection{A single party: the role of the leader}\label{sectIII1}
{
After this general introduction, we want to analyse next how the choice of a single agent is influenced  by its own party, and from its leader in particular. Assuming that all the members of the party are indistinguishable, it is natural to focus on a single member, $\Lc$, so to that the dimension of the problem reduces, and the relevant Hilbert space is just  $\Hil=\mathbb{C}^2$. The vector representing $\Lc$'s choice
is simply $\psi=\alpha |e_0\rangle +\beta |e_1\rangle$, as we already discussed before.
The only effects we consider here are the free will (or its uncertainty) of $\Lc$ and the  leader's influence.}

In this case, the Hamiltonian governing the various processes occurring for $\Lc$ is assumed to be
$$
H=H_{f}+H_{v},
$$
\be\label{hami}
H_{f}=\omega \hat a^\dagger \hat a,\qquad H_{v} =\lambda(\hat a^\dagger+\hat a),
\en
where $\omega,\lambda$ are non negative parameters. Here we have introduced the (fermionic) ladder operators $\hat a$ and $\hat a^\dagger$. These operators are very well known and used originally in quantum mechanics, see \cite{rom,mer} for instance, but then adopted also in other contexts, \cite{bagbook,bagbook2}. For our purposes, it is sufficient to say that these operators are defined on the o.n. basis $\{e_0,e_1\}$ of $\mathbb{C}^2$ as follows:
$$
\hat a e_0=0, \quad \hat a e_1=e_0, \qquad \hat a^\dagger e_0=e_1, \quad \hat a^\dagger e_1=0.
$$
They satisfy the canonical anti-commutation relations (CAR) $\{\hat a,\hat a^\dagger\}=\hat a\,\hat a^\dagger+\hat a^\dagger\,\hat a=\mathcal{I}_2$ and $\hat a^2=0$.

As widely discussed in the literature, see \cite{bagbook,bagbook2} for an overview, the term $H_f$ is responsible of an inertial mechanism which somehow forces $\Lc$ to maintain its initial choice, whereas $H_v$ works in the opposite way by inducing some change in $\Lc$ while forming its final decision. This can be  understood as follows: suppose that $\Lc$ is described, at $t=0$, by the vector $\psi=e_j$. Then $\psi$ is an eigenstate of $H_f$, so that $\psi$ is not modified when acting on it with $H_f$. On the other hand, suppose that $\psi=e_0$. Then, using (\ref{hami}), we see that $H_v\psi=\lambda e_1$, while, if  $\psi=e_1$, we find that $H_v\psi=\lambda e_0$: the action of $H_v$ on $\psi$ changes drastically the state of the system. Notice that, so far, there is no reason for $\Lc$ to change his status toward "no" or "yes". In other words, there is no reason for $\Lc$ to move from its original state $\psi$ to a {\em new} state $\psi_{new}$ which is either $e_1$ or $e_0$ (or close to them). The way this task can be achieved is by introducing in the model  the two Lindblad operators
\be
L_{\A,1}=\tau_1\hat a,\qquad L_{\A,2}=\tau_2\hat a^\dagger,
\label{add1}\en
which represent the action of $\Lc$'s leader on $\Lc$ itself. In particular, under the action of $L_{\A,1}$, $\Lc$ is  influenced by its leader to vote "yes", since it is {\em forced} to the vector $e_0$, whereas $L_{\A,2}$ does the opposite forcing to the vote "no", i.e., to the vector $e_1$: the real parameters $\tau_1$ and $\tau_2$ fix the strengths of these actions, and having $\tau_1$ and $\tau_2$ both non zero can be seen as the simultaneous presence of two different leaders of the same party proposing the two opposite final choice "yes" and "no" respectively. Of course, the more influent is the leader, the higher the value of its parameter $\tau_j$.
Looking at the action of the operators on a pure state identified by
$|\psi\rangle=\alpha |e_0\rangle+\beta |e_1\rangle$, for a small time $dt$, according to \eqref{linjumps} the pure state becomes the mixed state
\bea\label{add3}
\rho(dt)=p_A\mathcal{ \tilde A}|\psi\rangle \langle\psi|\mathcal{ \tilde A}^\dagger+p_{B_1}|e_0\rangle \langle e_0|+p_{B_2}|e_1\rangle \langle e_1|
\ena
with 
$$
\mathcal{\tilde A}=\frac{1}{\sqrt{p_A}}\left(\1-\frac{dt}{2}\left(|\alpha\tau_2|^2|e_0\rangle \langle e_0|+|\beta\tau_1|^2|e_1\rangle \langle e_1|\right)\right)
$$
and 
$$
p_A=(1-p_{B_1}-p_{B_2}),\qquad p_{B_1}=dt|\beta\tau_1|^2,\qquad p_{B_2}=dt|\alpha\tau_2|^2.
$$
Hence there is a chance that the  state is evolved in $|e_0\rangle$ (vote "yes") with probability $p_{B_1}$, in $|e_1\rangle$ (vote "no") with probability $p_{B_2}$, otherwise it follows the continuous drift-type evolution. It is expected that as time passes the mixture of the states becomes a relevant phenomenon.

%
%
%
%
%

\begin{figure}[!ht]
	\begin{center}		
		\hspace*{-0.55cm}\subfigure[]{\includegraphics[width=8.3cm]{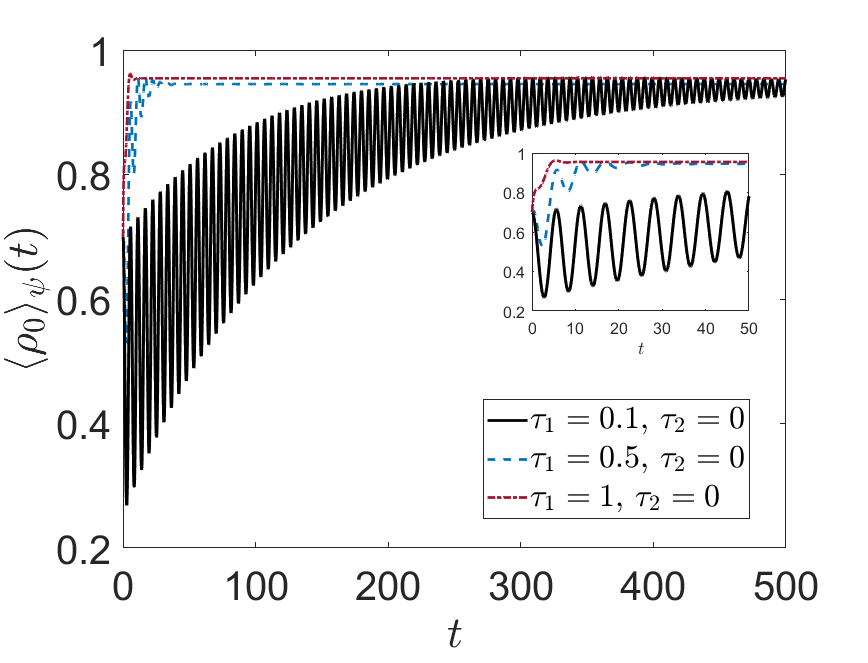}}	
		\hspace*{-0.55cm}\subfigure[]{\includegraphics[width=8.3cm]{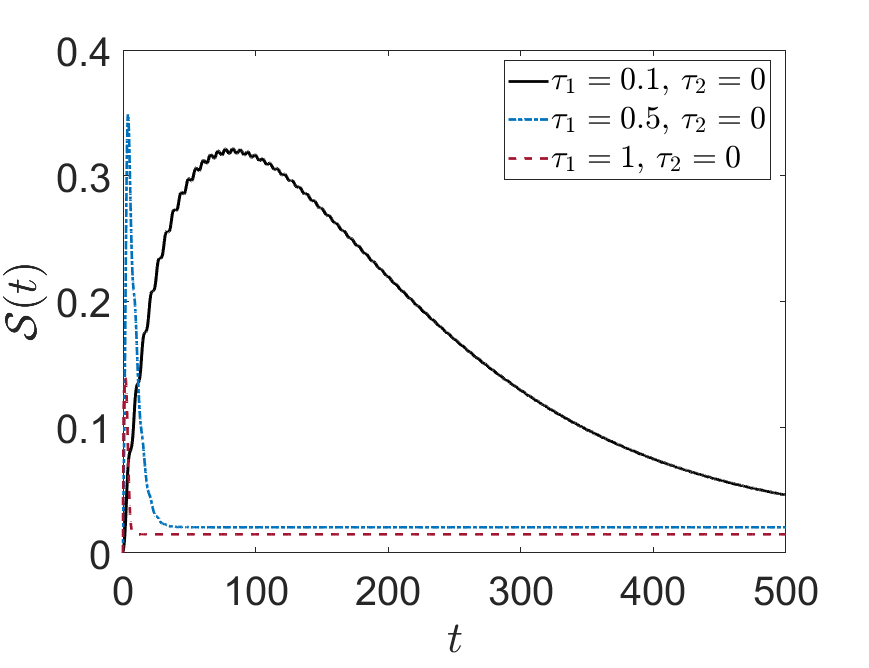}}
	\end{center}	
	\caption{(a) The time evolution of the mean value $\langle\rho_{0}\rangle_{\psi}(t)$ for different values of $\tau_1$ and with $\tau_2=0$. Other parameters: $\omega=1,\lambda=0.25$. The initial condition is $|\psi\rangle =\sqrt{0.7}|e_0\rangle+\sqrt{0.3}|e_1\rangle$. In the small inset the time evolution for early times. (b) The time evolution of the entropy $\mathcal{S}(t)$ for the same parameters and initial condition.} 
	\label{fig:tau1}
\end{figure}

\begin{figure}[!ht]
	\begin{center}		
		\hspace*{-0.55cm}\subfigure[]{\includegraphics[width=8.3cm]{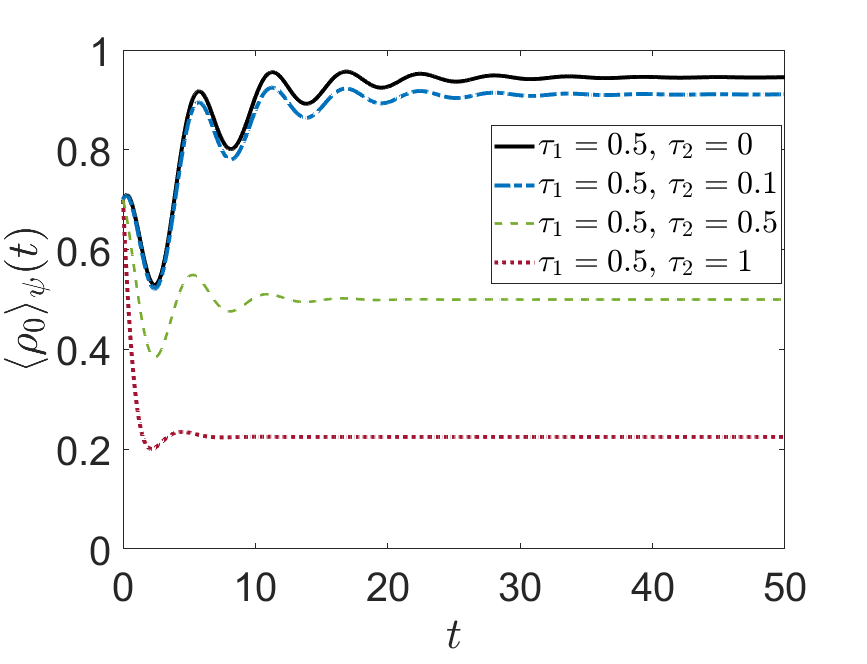}}	
		\hspace*{-0.55cm}\subfigure[]{\includegraphics[width=8.3cm]{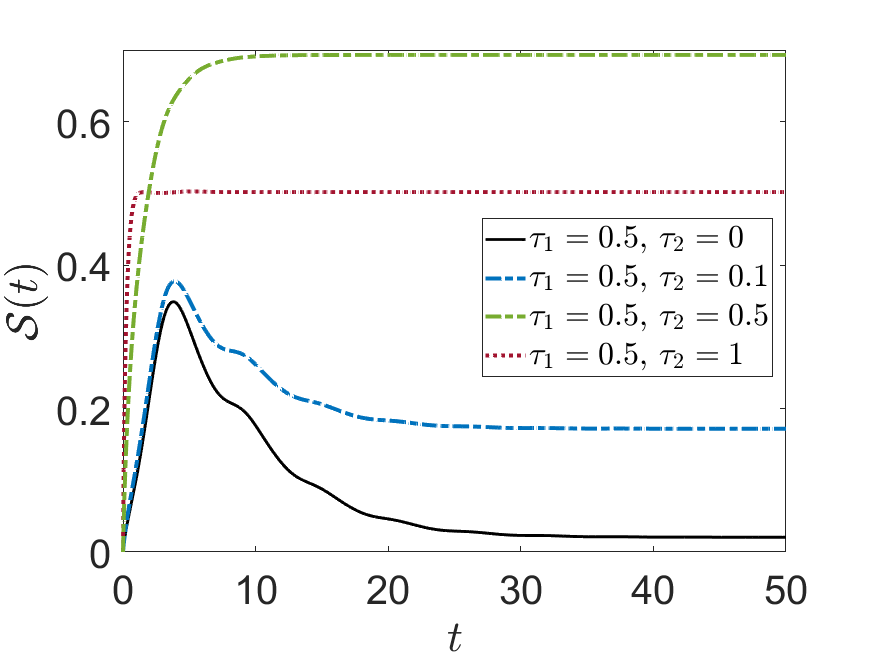}}
	\end{center}
	\caption{(a) The time evolution of the mean value $\langle\rho_{0}\rangle_{\psi}(t)$  for different values of $\tau_2$ and with $\tau_1=0.5$. Other parameters: $\omega=1,\lambda=0.25$. The initial condition is  $|\psi\rangle =\sqrt{0.7}|e_0\rangle+\sqrt{0.3}|e_1\rangle$. (b) The time evolution of the entropy $\mathcal{S}(t)$ for the same parameters and initial condition. }
	\label{fig:tau2}
\end{figure}

Some numerical simulations for different values of $\tau_1$ and $\tau_2$ are shown in Figures \ref{fig:tau1}(a)-(b) and \ref{fig:tau2}(a)-(b), where the mean values $\langle \rho_0\rangle_{\psi}$, defined in \eqref{meanrhob}, and the entropy $S$, defined in \eqref{entro}, are shown.
Considering the case $\tau_1\neq0,\tau_2=0$, that  represents the situation where $\Lc$ is influenced by Alice to vote "yes", we reach an equilibrium that, depending also on the balance with the other  contribution in \eqref{eq:lineq}, tends faster to 1 as $\tau_1$ increases. We see that the whole dynamics behaves as one would expect, given that the final choice to vote "yes" turns out to be highly probable\footnote{Even if the long time value of $\langle \rho_0\rangle_{\psi}$ is not exactly one, it appears to be {\em almost one}. This is a sort of uncertainty in our model, which replace the free will in \cite{andro}.}.

We  notice that, for moderate low values of $\tau_1$, see the cases $\tau_1=0.1$ and $\tau_1=0.5$, strong amplitude oscillations are visible in the early-mid phase and tend to be damped for later times. {They are consequences of the member's indecision mechanism due to the Hamiltonian term $H_v$ which, as observed in other contexts (\cite{bagbook,bagbook2}), is the main responsible of the oscillatory behaviour in the mean values of the density or number operators.
We should consider that, in a real situation, there could be also (few) members of $\Pc_{\A}$ reaching a different final choice and therefore voting "no". Hence, from a pure quantum interpretation, the final state is a mixture of states most of them representing the vote "yes" and few others the vote "no". The evolution of the entropy $S$ can be seen in this sense as a possible measure of this mixture. In Figures \ref{fig:tau1}(b) we observe that  $S(t)$ is always strictly positive, indicative of the presence of a mixture of states, and showing the presence of a main peak in time followed by a rapid decreasing up to some equilibrium value. This value is reached more rapidly for increasing values of $\tau_1$. We can justify this behaviour by imagining that in a first phase the influence of $\cal{L}$'s leader due to the Lindblad operator $L_{\A,1}$ strongly modifies the member's state of mind, also because of the presence of $H_v$. However, the higher is the strength of $L_{\A,1}$, i.e. the value of $\tau_1$, the more rapidly this condition moves toward an equilibrium. The fact that the final equilibrium value increases for lower $\tau_1$ can be interpreted as the lower influence of $\cal{L}$'s leader and hence to a richer mixture of states related in principle to the presence of more  members of the party voting "no" (which, however, remain a minority with respect to those voting "yes")\footnote{This interpretation is based on the assumption that all the members of the various parties are indistinguishable, so that each member of $\Pc_{\A}$ behaves as $\Lc$.}.

\vspace{2mm}

Adding the Lindblad operator $L_{\A,2}$ creates a {\em richer} dynamics, as shown in Figures  \ref{fig:tau2}(a)-(b). The presence of two Lindblad operators, inducing opposite effects, can be seen as a situation in which two different leaders of the same party influence the final choice of $\Lc$. As expected when $\tau_1>\tau_2$ the final mean value  $\langle \rho_0\rangle_{\psi}$ is closer to 1 rather than to 0, and it approaches 1 more and more as the difference $\tau_1-\tau_2$ increases. A perfect equilibrium is reached when $\tau_1=\tau_2$, and in this case the final value of $\langle \rho_0\rangle_{\psi}$ is equal to 0.5. Finally, for $\tau_2>\tau_1$,  $\langle \rho_0\rangle_{\psi}$ is closer to 0, regardless its initial value. The various mechanisms that could lead to this dynamics can be straightforwardly deduced by the previous discussion made on the case $\tau_2=0$. Concerning the time evolution of the entropy $S$ we can observe that the case $\tau_1=\tau_2$ can be considered critical in the sense that the equilibrium value is $\log (2)\approx 0.693$ which is the maximum admissible value for $S$ (we recall that $S$ is bounded by the value $\log{(d)}$, being $d$ the dimension of the Hilbert space in this case). This case represents a situation of uncertainty in which no clear final choice is achieved, and the state is a {\em perfect mixture} of $\rho_0$ and $\rho_1$.
 The other cases show also that adding a second Lindblad operators induces a stronger mixture as compared to the case where $\tau_2=0$: two competing leaders of a single party create more uncertainty!}

\subsection{Two parties: the dynamics of alliance}
In this section we want to model a situation in which the members of  the first party $\mathcal P_\A$ interact with the members of the second party $\mathcal P_\B$, assuming however that only the members of $\mathcal{P}_\A$ receive specific indications on the option they should vote (specifically the option "yes"). It is like if Bob's influence on his own party is very low, if not completely negligible: this is what is usually called {\em lack of leadership}. We will consider the case in which Bob influences the various $p_\B(j)$ later on. Our goal here is to derive the proper model and operators to describe: i) a collaborative-like  attitude of the parties, that is the second members of $\mathcal P_\B$ are inclined to vote for the same option as the members of $\mathcal P_\A$;
ii) a conflictual dynamics in which the two parties move in different directions (one votes "yes" while the other votes  "no"). 

More in details, we suppose here that only the Lindblad operator $L_{\A,1}$ in (\ref{add1}) acts by forcing  the members of $\mathcal{P}_\A$ to vote "yes", and that the behaviour of the members of $\mathcal{P}_\B$ is only dictated by their own interactions with $\mathcal P_\A$. It follows that the Lindblad operator is simply $L_{\A,1}=\tau_1\hat a_1$, whereas the Hamiltonian ruling the interactions between the members can be assumed to be 
$$
H=H_f+H_v+H_{c}+H_{nc},
$$
where $H_f,H_{v}$ here extend those given in the previous section, while $H_c,\,H_{nc}$ are operators describing the new collaborative or conflictual (non collaborative) dynamics.
To be  specific,
\bea
H_f&=&\omega_1 \hat{a}_1^\dagger \hat{a}_1+\omega_2 \hat{a}_2^\dagger \hat{a}_2,\,\label{216a}\\
H_v&=&\lambda_1 \left(\hat{a}_1^\dagger +\hat{a}_1\right)+\lambda_2\left(\hat{a}_2^\dagger+ \hat{a}_2\right),\,\label{216b}
\ena
with $\omega_{1,2}\geq0,\lambda_{1,2}\geq0$,
while  the new  operators are
\bea
H_{c}=\gamma_{c} \left(\hat{a}_{1}^\dagger\hat{a}_{2}+\hat{a}_{2}^\dagger\hat{a}_{1}\right),\qquad H_{nc}=\gamma_{nc} \left(\hat{a}_{1}^\dagger\hat{a}^\dagger_{2}+\hat{a}_{2}\hat{a}_{1}\right).
\ena
Here our interest is focused on just two legislators, $\Lc_1$ and $\Lc_2$, as representants of Alice's and Bob's parties, assuming as in Section \ref{sectIII1} that all the members of a given party share a similar attitude towards $\Lambda$. The operators $\hat a_j$ and $\hat a_j^\dagger$ obey the following two-dimensional CAR:
\be
\{\hat a_j,\hat a_k^\dagger\}=\delta_{j,k}\mathcal{I}_4, \qquad \{\hat a_j,\hat a_k\}=0,
\label{add2}\en
$j,k=1,2$. Here $\mathcal{I}_4$ is the identity operator on $\mathbb{C}^4$.

The motivation that leads to these terms in the Hamiltonian is based on the way in which the Lindblad operator acts on the system. In fact, in view of what we have seen before, $L_{\A,1}$ drives a generic vector $|\tilde \psi\rangle=\sum_{j,k=0,1}\alpha_{j,k}|e_{j,k}\rangle$ into a new vector where the components proportional to $\alpha_{1,0}$ and $\alpha_{1,1}$ {\em tend} to disappear\footnote{In fact, the action of $L_{\A,1}$ makes these terms in $|\tilde \psi\rangle$ disappear. However, the simultaneous effect of $H$, partly restores them.}.

Given that, it is interesting to describe how, at least heuristically, the combined action of $H$ and $L_{\A,1}$ works on the members of $\mathcal{P}_\mathcal{B}$  supposing that the initial configuration is represented by the pure state $|\psi(0)\rangle$ (with its related density operator $\rho=|\psi(0)\rangle \langle \psi(0)|$) given by 
$$
|\psi(0)\rangle=\sum_{j,k=0,1}\alpha_{j,k}|e_{j,k}\rangle,\,\textrm{ with } \sum_{j,k=0,1}|\alpha_{j,k}|^2=1.
$$
Since the behaviour of the members of $\mathcal{P}_\mathcal{B}$ is not directly modified by $L_{\A,1}$, and observing that $H_f, H_{v}$ are not responsible for the interaction, we start describing naively what happens if we focus for small times on a branch which is firstly affected by the action of $H_c$ and then by $L_{\A,1}$. In particular, we have
\bea
H_c|\psi(0) \rangle=\gamma_c\left(\alpha_{1,0}|e_{0,1}\rangle+\alpha_{0,1}|e_{1,0}\rangle\right),
\ena
since all the other terms of $|\psi(0) \rangle$ are annihilated by the action of $H_c$. Hence,  after a small time $dt$, we obtain the new vector (up to a suitable normalization)  
\bea
|\psi(dt)\rangle\approx|\psi(0) \rangle-i dt\,\gamma_c\left(\alpha_{1,0}|e_{0,1}\rangle+\alpha_{0,1}|e_{1,0}\rangle\right).\label{216}
\ena
This vector is different from $|\psi(0) \rangle$ if $\alpha_{1,0}$ or $\alpha_{0,1}$ are non zero.
Then, following the same ideas discussed in the previous sections, and considering the subsequent action of  $L_{\A,1}$ only, the obtained state 
is  induced to jump, with some non zero probability, to a state expressed by the density operator
$$
\mathcal{\tilde B}_1\, |\psi(dt)\rangle\langle\psi(dt)|\mathcal{\tilde B}_1^\dagger,
$$
where $\mathcal{\tilde B}_1=\frac{1}{\| \hat{a}_1|\psi\rangle\|}\hat{a}_1$ (see \eqref{Bj}).
It is clear that, in view of \eqref{216}, the state following this jump can only be of the form 
\bea
|\psi_{\textrm{jump}}\rangle=\sum_{k=0,1}\tilde\alpha_{1,k}|e_{0,k}\rangle\,-i\,dt\gamma_c\alpha_{0,1}|e_{0,0}\rangle,\label{j217}
\ena
where $\tilde\alpha_{1,k}=\alpha_{1,k}/|| \hat{a}_1|\psi\rangle||$
and where as usual the proper normalization should be inserted.
It is now clear that the second populations can have an excitement of the 0 level due to the appearance of the term $-i\,dt\gamma_c\alpha_{0,1}|e_{0,0}\rangle$:  the two parties are driven to the same decision. This perturbative analysis reflects our numerical results, as we will show next.

With similar computations, we can derive the approximated state obtained by the action of $H_{nc}$ first, and the jump induced by $L_{\A,1}$ after. In this case we get
\bea
|\psi_{\textrm{jump}}\rangle=\sum_{k=0,1}\tilde\alpha_{1,k}|e_{0,k}\rangle\,-i\,dt\gamma_{nc}\alpha_{0,0}|e_{0,1}\rangle,
\ena
which, when compared to \eqref{j217}, leads to an excitement of  $|e_{0,1}\rangle$, that is the state representing the two opposite members' decision.

\begin{figure}[!ht]
	\begin{center}		
		\hspace*{-0.4cm}\subfigure[]{\includegraphics[width=8.3cm]{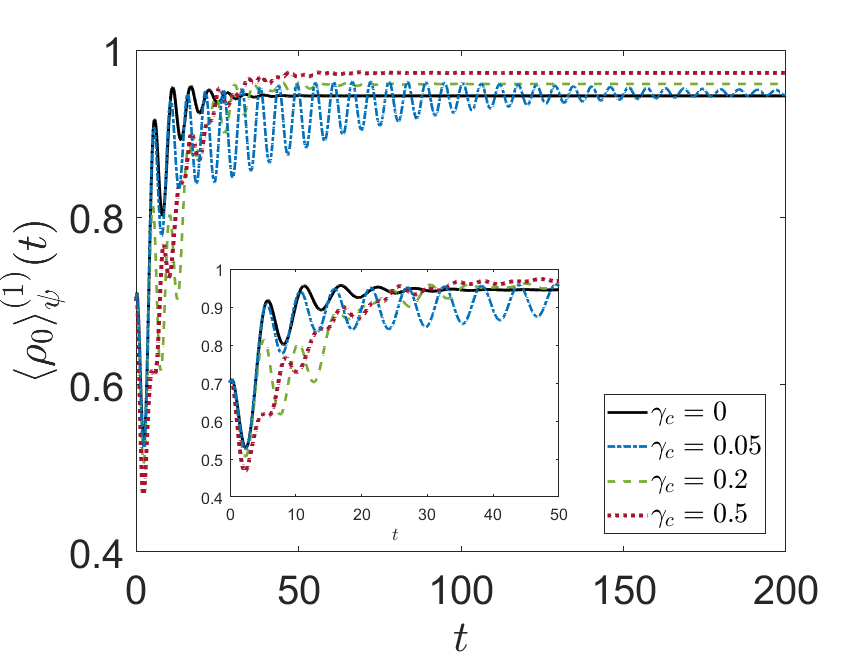}}	
		\hspace*{-0.4cm}\subfigure[]{\includegraphics[width=8.3cm]{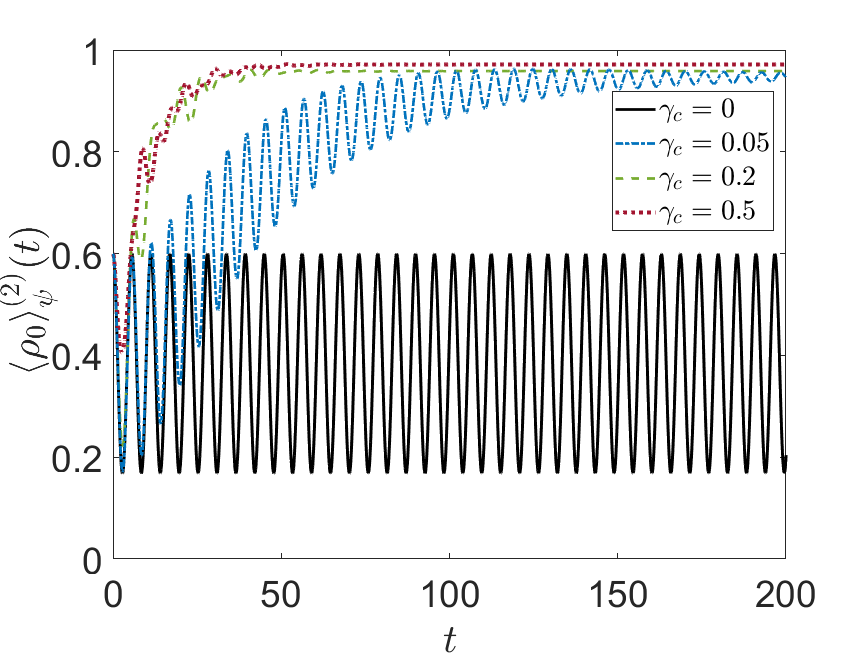}}
		\hspace*{-0.4cm}\subfigure[]{\includegraphics[width=8.3cm]{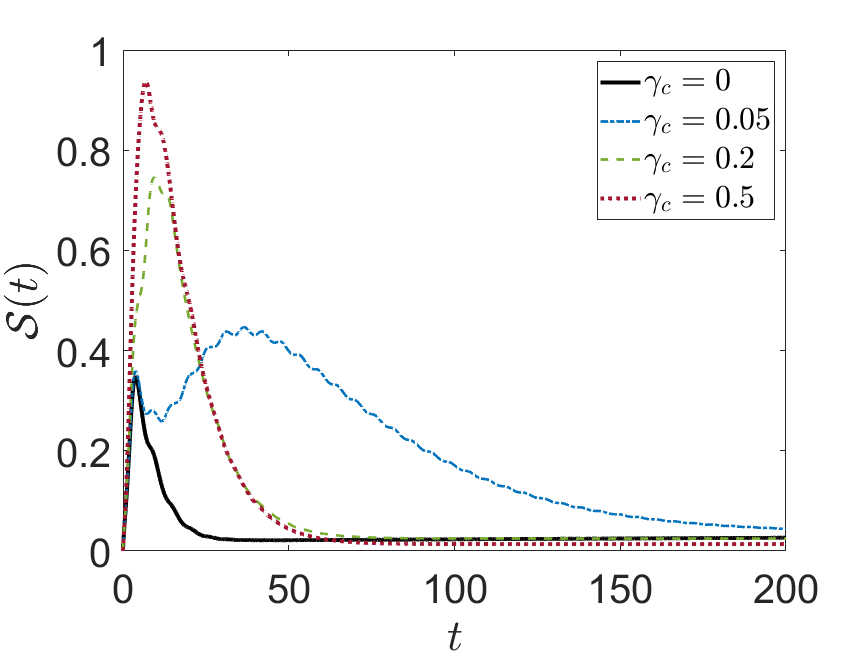}}
	\end{center}
	\caption{(a) The time evolution of the mean value $\langle\rho_{0}\rangle_{\psi}^{(1)}(t)$ for different values of $\gamma_c$ (cooperative attitudes). Other parameters: $\omega_1=\omega_2=1,\,\lambda_1=\lambda_2=0.25,\,\gamma_{nc}=0,\tau_1=0.5$. Initial state is such that $\langle\rho_{0}\rangle_{\psi}^{(1)}=0.6,\langle\rho_{0}\rangle_{\psi}^{(2)}=0.4$. (b) The time evolution of the mean  value $\langle\rho_{0}\rangle_{\psi}^{(2)}(t)$  for the same parameters and initial condition. (c) The time evolution of the entropy $\mathcal{S}(t)$ for the same parameters and initial condition.}
	\label{fig:coop}
\end{figure}

\begin{figure}[!ht]
	\begin{center}		
		\hspace*{-0.4cm}\subfigure[]{\includegraphics[width=8.3cm]{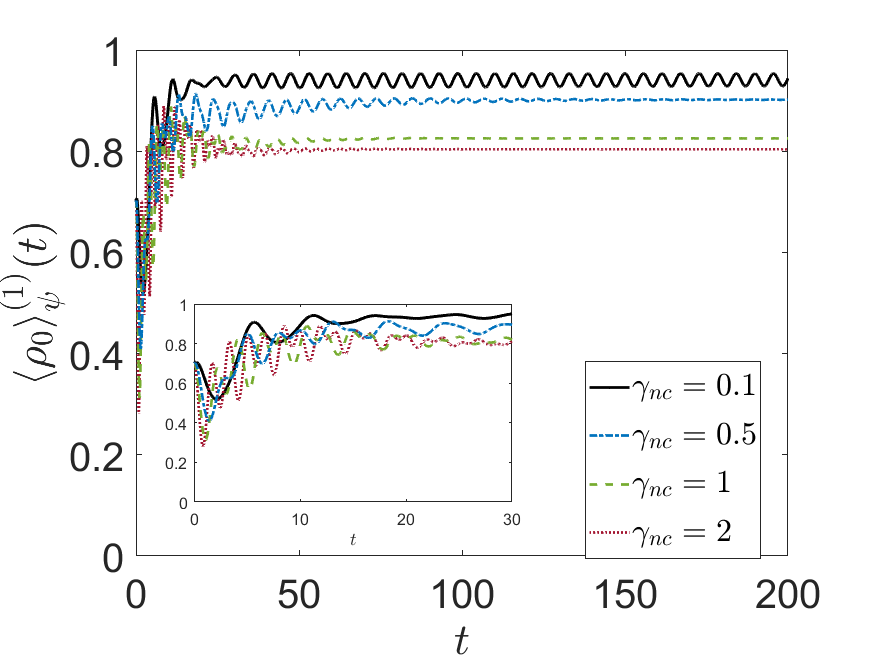}}	
		\hspace*{-0.4cm}\subfigure[]{\includegraphics[width=8.3cm]{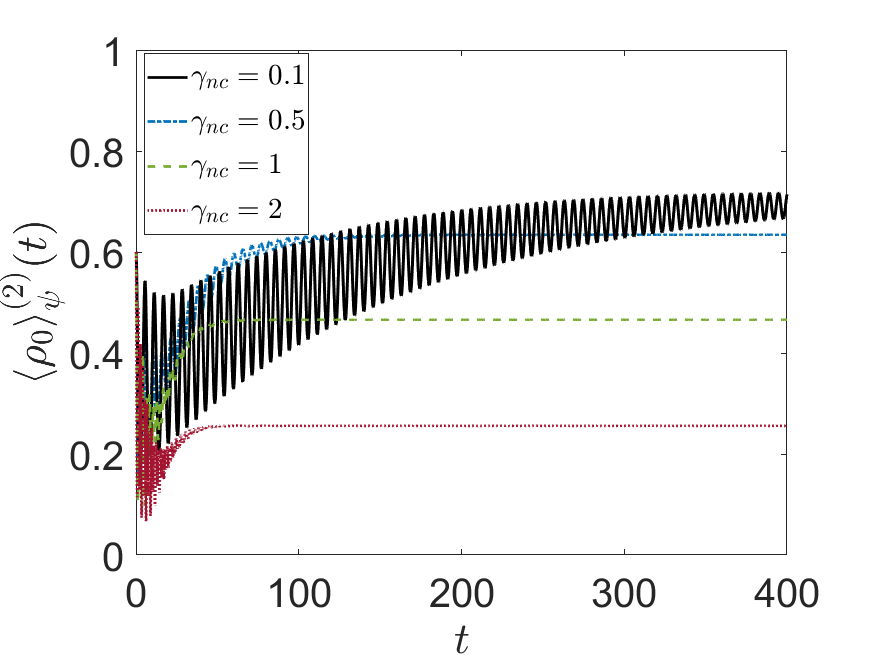}}
			\hspace*{-0.4cm}\subfigure[]{\includegraphics[width=8.3cm]{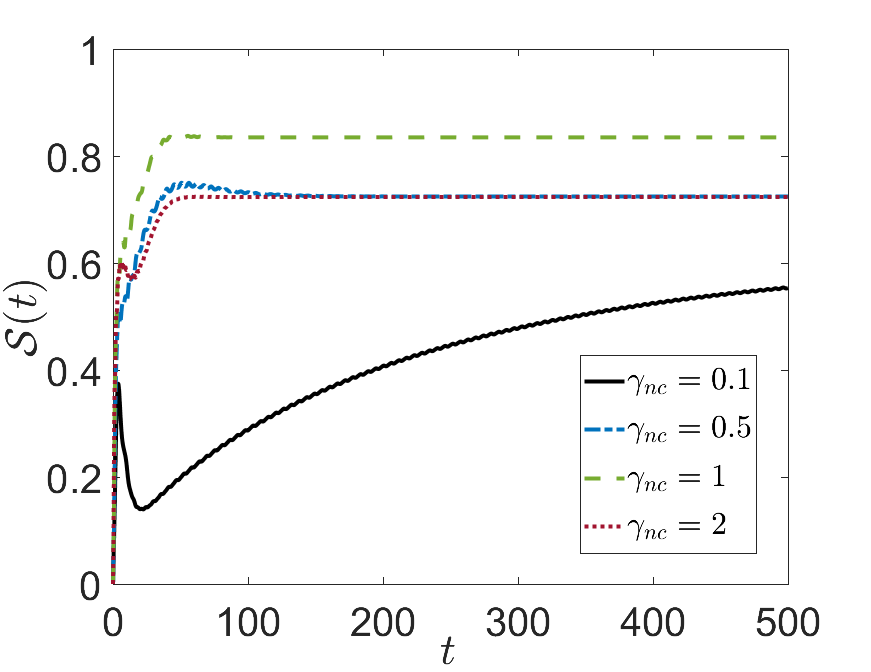}}
	\end{center}
	
	\caption{ The time evolution of the expected value $\langle\rho_{0}\rangle_{\psi}^{(1)}(t)$ for different values of $\gamma_{nc}$ (non cooperative attitudes). Other parameters: $\omega_1=\omega_2=1,\,\lambda_1=\lambda_2=0.25,\,\gamma_{c}=0,\tau_1=0.5$. Initial state is such that $\langle\rho_{0}\rangle_{\psi}^{(1)}=0.6,\langle\rho_{0}\rangle_{\psi}^{(2)}=0.4$. (b) The time evolution of the expected value $\langle\rho_{0}\rangle_{\psi}^{(2)}(t)$  for the same parameters and initial condition. (c) The time evolution of the entropy $\mathcal{S}(t)$ for the same parameters and initial condition.}
	\label{fig:contr}
\end{figure}

Numerical results confirming this kind of behavior are shown in Figure \ref{fig:coop} for the cooperative attitude, and in  Figure \ref{fig:contr} for the non cooperative one. Figures contain the time evolutions of the legislator's intention to vote "yes" for both parties, expressed by the mean values 
\bea
\langle\rho_{0}\rangle_{\psi}^{(1)}=\textrm{Tr}\left[\rho_\psi(\rho_0\otimes \mathcal{I}_2)\right],\qquad \langle\rho_{0}\rangle_{\psi}^{(2)}=\textrm{Tr}\left[\rho_\psi(\mathcal{I}_2\otimes \rho_0 )\right],\label{meanrho}
\ena
where $\rho_{0}=\begin{pmatrix}
1 & 0 \\
0 & 0 
\end{pmatrix},\,\mathcal{I}_2=\begin{pmatrix}
1 & 0 \\
0 & 1 
\end{pmatrix},$
and of the entropy $\mathcal{S}(t)$.

As we can see in Figures \ref{fig:coop}(a)-(b), for the cooperative case, increasing the value of $\gamma_c$ leads to a continuous growth of the asymptotic value of $\langle\rho_{0}\rangle_{\psi}^{(2)}(t)$, showing that the action of $H_c$ is responsible of the common attitude of the members of $\mathcal{P}_\A$ and $\mathcal{P}_\B$. We also notice that, again for increasing $\gamma_c$, also $\langle\rho_{0}\rangle_{\psi}^{(1)}(t)$ slightly increases its asymptotic value: we can speculate on this effect by supposing that the members of $\mathcal{P}_\A$ reinforce their attitude in voting yes when they interact and influence the members of $\mathcal{P}_\B$ in doing the same. This is also supported by \eqref{j217} according to which there is a jump proportional to $\gamma_c$ toward the state $|e_{0,0}\rangle$, that is all members vote "yes". Concerning the measure of the entropy $\mathcal{S}(t)$, Figure \ref{fig:coop}(c), the initial phase (up to $t\approx50$) is characterized by an overall growth of $\Sc(t)$, and then by a decreasing behaviour and convergence towards an asymptotic value which in general decreases faster for increasing $\gamma_c$. The peaks in $\mathcal{S}(t)$ are higher for larger values of $\gamma_c$. This could be interpreted by the fact that the cooperative dynamics induced by $H_c$ creates, together with the action of $L_{\A,1}$, a rapid mixture of states representing the same final decision taken by the parties, and the rapid decay can be seen as an immediate settlement to the  asymptotic value. 

The non-cooperative case is shown in Figures \ref{fig:contr}, where the various time evolutions are shown by changing the key parameter $\gamma_{nc}$ which tunes the strength of $H_{nc}$. As expected, and predicted by our perturbative approach,  increasing the effect of $H_{nc}$ leads to two opposite final choices taken by the members of $\cal{P}_{\cal{A}}$ and $\cal{P}_{\cal{B}}$, even if the behaviour of these latter is not as sharp as that of $\Pc_\A$. It is interesting to note that also the entropy $\Sc(t)$ attains its equilibrium to larger values than those obtained in the cooperative case; this can be explained by the fact that the action of the Lindbladian operator $L_{\A,1}$ together with the action of $H$, induces a strong mixture of states increased by the non collaborative effect.

\subsection{The effects of Bob and $\Pc_\M$}

	We conclude the analysis of our quantum parliament $\pi$ by briefly considering the inclusion in our model of a third party $\Pc_{\M}$ that does not follow a leader's influence, and the case in which Bob has some (weak) influence on $\Pc_\B$.
	{We shall perform two different kind of experiments. In the first one, we suppose that only $\Pc_\A$ is influenced by its leader, whereas the members of  $\Pc_\B$ and $\Pc_\M$ are left to interact with the members of $\Pc_\A$ but they both have no strong leader to follow. In the second experiment we suppose that $\Pc_\B$ has a leader, Bob,  {\em weak} compared to Alice. The presence of a third party in our model is interesting because  allows us to consider non-trivial cubic terms in the Hamiltonian of $\pi$, see (\ref{hcubica}) below.}
	
	 For the first experiment, we use the same hypothesis of the previous sections,  assuming that  the only Lindblad operator is $L_{\A,1}=\tau_1\hat a_1$.
	In this case, adding a third agent to our system, the  Hilbert space of the micro-system becomes $\Hil=\mathbb{C}^8$, and the vector representing the member's choice is given by $| \psi\rangle=\sum_{j,k,l=0,1}\alpha_{j,k,l}|e_{j,k,l}\rangle$ with $\sum_{j,k,l=0,1}|\alpha_{j,k,l}|^2=1$.
    Concerning the Hamiltonian ruling the behaviour of the three parties, it is natural to consider the following one:
    \bea
H&=&H_{f}+H_{v}+H_{i},\label{H3}
\ena
where, similarly to \eqref{216a}-\eqref{216b} we have

\bea
H_f&=&\sum_{k=1,2,3}\omega_k \hat{a}_k^\dagger \hat{a}_k,\\
H_v&=&\sum_{k=1,2,3}\lambda_k \left(\hat{a}_k^\dagger +\hat{a}_k\right)
\ena
with $\omega_{1,2,3}\geq0,\lambda_{1,2,3}\geq0$,
and where the interaction term $H_{int}$
is assumed to be
\bea\label{hcubica}
 H_{int}=\gamma_{1}\,\hat{a}_1^\dagger\hat{a}_2\hat{a}_3+
\gamma_{2}\,\hat{a}_1^\dagger\hat{a}_2^\dagger\hat{a}_3+
\gamma_{3}\,\hat{a}_1^\dagger\hat{a}_2\hat{a}_3^\dagger+
\gamma_{4}\,\hat{a}_1^\dagger\hat{a}_2^\dagger\hat{a}_3^\dagger+
\text{h.c.}
\ena
which contains all the possible cubic terms related to the ways the three parties could mutually interact, and where $\gamma_{k}\geq0,\,k=1,2,3,4$. Here h.c. stands for hermitian conjugate. This is needed if we require $H=H^\dagger$. The CAR for the operators involved extend those in (\ref{add2}) to three dimensions.
To clarify the effect of each term in $ H_{int}$, in our numerical simulations we always assume that only one of the parameters  $\gamma_k$ is different from zero. Of course, one can easily relax this assumption to create more complex dynamics considering various terms in $ H_{int}$ acting simultaneously.  

\begin{figure}[!ht]
	\begin{center}	
		\hspace*{-0.4cm}\subfigure[]{\includegraphics[width=8.3cm]{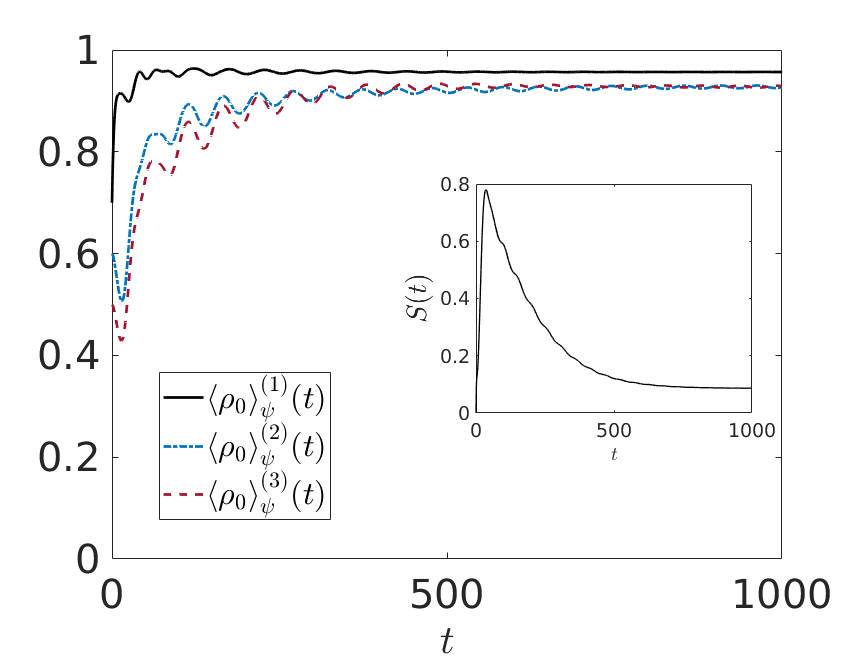}}	
		\hspace*{-0.4cm}\subfigure[]{\includegraphics[width=8.3cm]{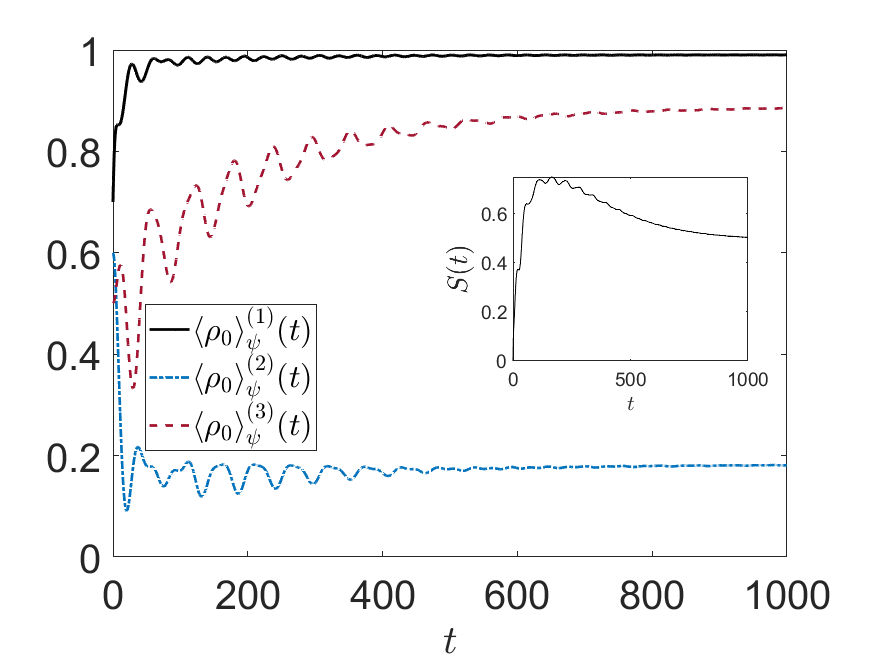}}
		\hspace*{-0.4cm}\subfigure[]{\includegraphics[width=8.3cm]{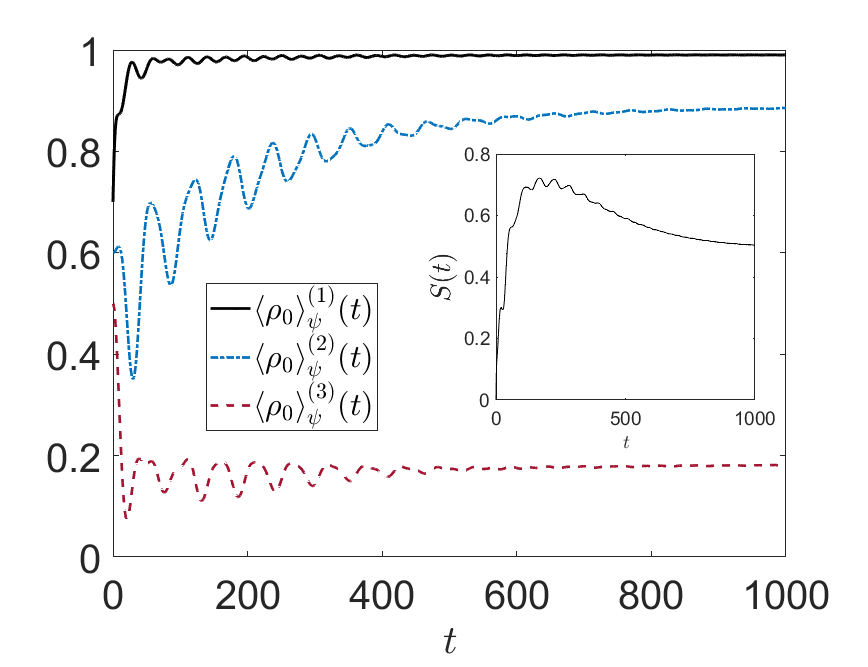}}	
		\hspace*{-0.4cm}\subfigure[]{\includegraphics[width=8.3cm]{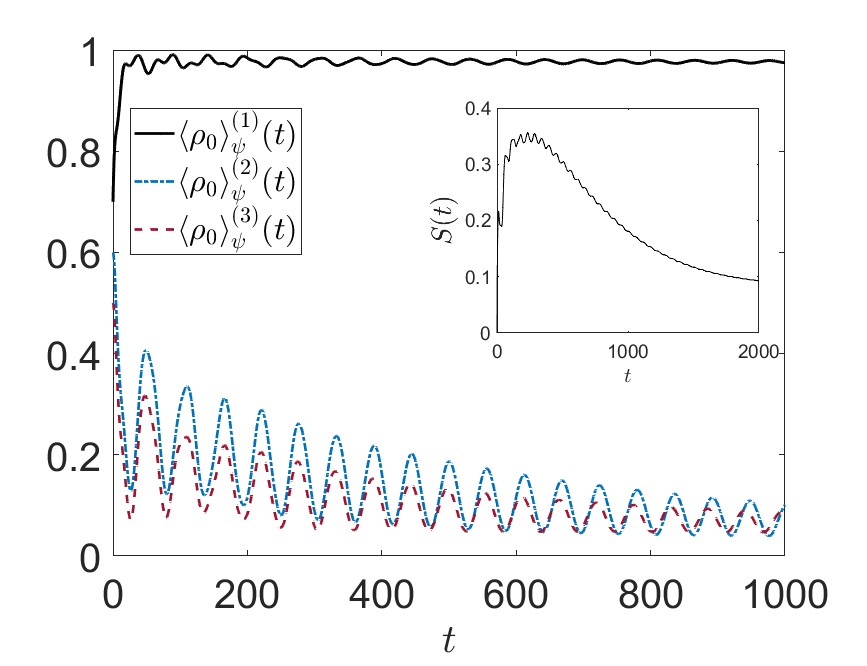}}
	\end{center}
	
	\caption{(a) The time evolution of the expected value $\langle \rho_0\rangle_{\psi}^{(1,2,3)}(t)$  for the three parties model. Non zero parameters are: $\omega_1=\omega_2=\omega_3=0.1,\,\lambda_1=\lambda_2=\lambda_3=0.025,\,\gamma_{1}=1,\tau_1=0.5$. Initial state is such that $\langle \rho_0\rangle_{\psi}^{(1)}(0)=0.7,\langle \rho_0\rangle_{\psi}^{(2)}(0)=0.6,\langle \rho_0\rangle_{\psi}^{(3)}(0)=0.5$. In the inset the time evolution of the entropy $S$. (b) Same plots as in (a) with $\gamma_2=1,\gamma_1=0$, and the same other parameters and initial conditions. (c) Same plots as in (a) with $\gamma_3=1,\gamma_1=0$, and the same other parameters and initial conditions. (d) Same plots as in (a) with $\gamma_4=1,\gamma_1=0$, and the same other parameters and initial conditions.
	}
	\label{fig:3p}
\end{figure}

Some numerical results for different cases of $ H_{int}$ are shown in Figures
\ref{fig:3p}, where the 4 possible triple interactions are considered. Here we plot the functions
\be
\begin{cases}
&\langle\rho_{0}\rangle_{\psi}^{(1)}=\textrm{Tr}\left[\rho_\psi(\rho_0\otimes \mathcal{I}_2\otimes \mathcal{I}_2)\right],\\ &\langle\rho_{0}\rangle_{\psi}^{(2)}=\textrm{Tr}\left[\rho_\psi(\mathcal{I}_2\otimes \rho_0\otimes \mathcal{I}_2 )\right],\\
&\langle\rho_{0}\rangle_{\psi}^{(3)}=\textrm{Tr}\left[\rho_\psi(\mathcal{I}_2\otimes\mathcal{I}_2\otimes \rho_0 )\right],\label{meanrho2}
\end{cases}
\en
which extend those in (\ref{meanrho}), and are a measure of the members' will to vote "yes".

We notice that most of the results can be understood by following the same analysis, based on the perturbative approach, proposed in the previous section for the single party and the two-parties cases. Hence, for instance,  the case $ H_{int}=\gamma_1\hat a_1^\dagger\hat a_2\hat a_3+\text{h.c}$ corresponds to a dynamics in which all the members of the parties tend to vote "yes" (panel (a)), whereas the case $ H_{int}=\gamma_4\hat a_1^\dagger\hat a_2^\dagger\hat a_3^\dagger+\text{h.c}$ implies that the 
members of $\cal{P}_{\cal{A}}$  vote "yes", whereas the others vote "no" (panel (d)). They suggest a {\em globally collaborative} and a {\em partly non collaborative} behaviour. Other cases in the other panels follow straightforwardly.

\vspace{2mm}

{\bf Remark:--} It is worth observing that the perturbative approach considered before for one or two parties, is much less clear in this case, with three different parties. The presence of too many agents makes the global dynamics much more complicated, and the perturbation expansion in, e.g., (\ref{add3}) is less explicative of the full dynamics, especially when the strength of $L_{\A,1}$ is lower than that of $H$, that is $\tau_1$ is smaller as compared to the other parameters of the model.

\begin{figure}[!ht]
	\begin{center}	
		\hspace*{-0.4cm}\subfigure[]{\includegraphics[width=8.3cm]{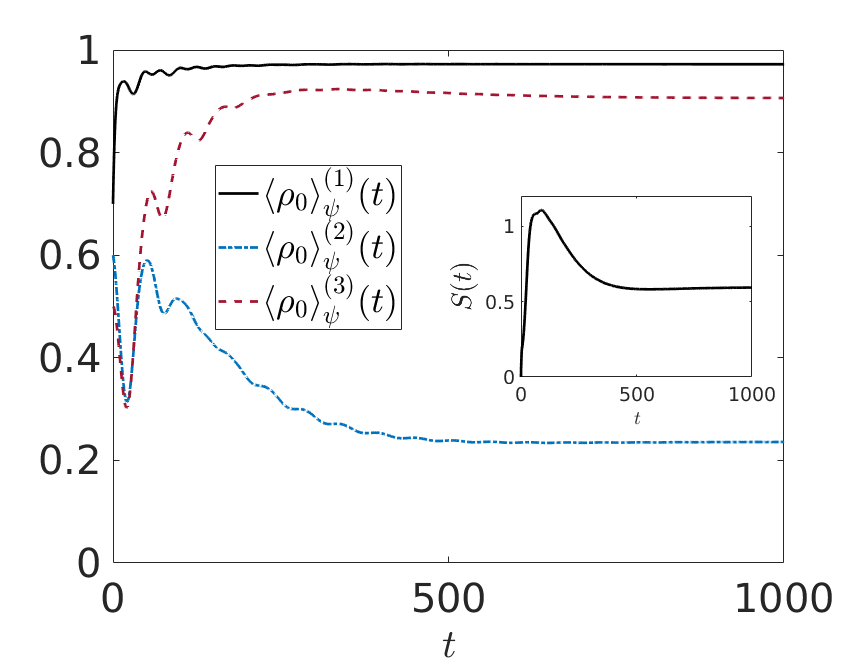}}	
		\hspace*{-0.4cm}\subfigure[]{\includegraphics[width=8.3cm]{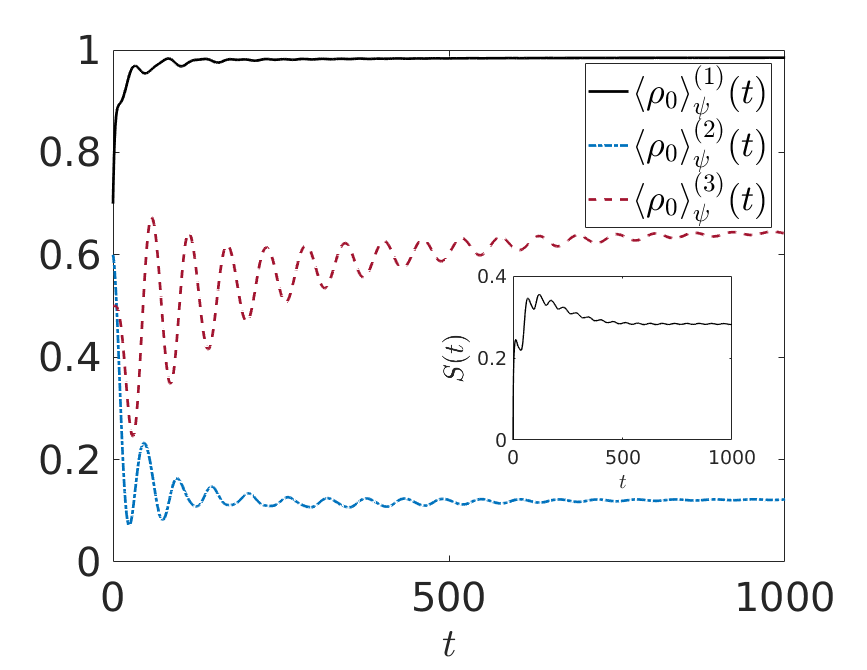}}
		\hspace*{-0.4cm}\subfigure[]{\includegraphics[width=8.3cm]{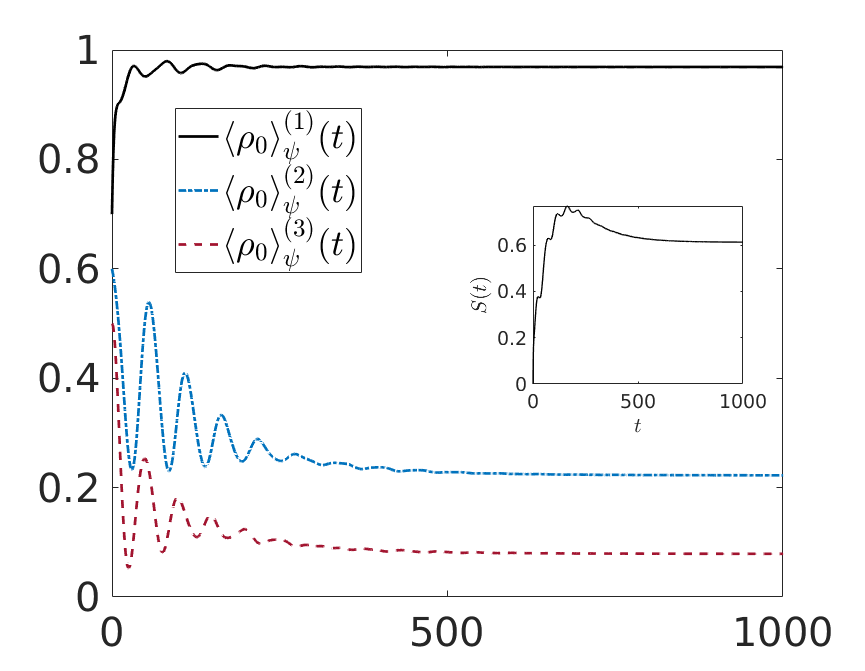}}	
		\hspace*{-0.4cm}\subfigure[]{\includegraphics[width=8.3cm]{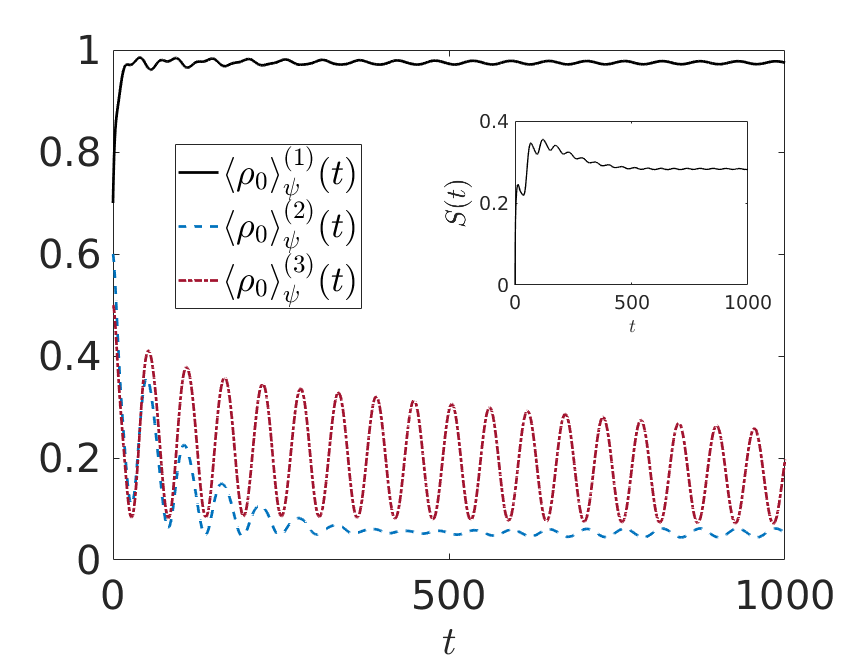}}
	\end{center}
	
	\vspace*{-0.7cm}\caption{(a) The time evolution of the expected value $\langle \rho_0\rangle_{\psi}^{(1,2,3)}(t)$  for the three parties model with the add of a second Lindblad operator $\kappa \hat{a}^\dagger_2$. Non zero parameters are: $\kappa=0.1$, $\omega_1=\omega_2=\omega_3=0.1,\,\lambda_1=\lambda_2=\lambda_3=0.025,\,\gamma_{1}=1,\tau_1=0.5$. Initial state is such that $\langle \rho_0\rangle_{\psi}^{(1)}(0)=0.7,\langle \rho_0\rangle_{\psi}^{(2)}(0)=0.6,\langle \rho_0\rangle_{\psi}^{(3)}(0)=0.5$. In the inset the time evolution of the entropy $S$. (b) Same plots as in (a) with $\gamma_2=1,\gamma_1=0$, and the same other parameters and initial conditions. (c) Same plots as in (a) with $\gamma_3=1,\gamma_1=0$, and the same other parameters and initial conditions. (d) Same plots as in (a) with $\gamma_4=1,\gamma_1=0$, and the same other parameters and initial conditions.
	}
	\label{fig:3p2_2L}
\end{figure}

\vspace{2mm}
The second set of experiments is performed by adding a second Lindblad operator that forces the members of $\Pc_\B$ to vote "no", opposite to the choice of $\Pc_\A$. In particular the Lindblad operator governing this mechanism is  
\bea
L_{\B,1}=\kappa \hat{a}^\dagger_2,
\ena
with $\kappa>0$. According to the analysis performed in the previous section, this term forces the generic member of $\Pc_\B$ to jump in a state representing the vote "no". We also suppose that the Hamiltonian is same defined in \eqref{H3}. Numerical results are shown in Figures \ref{fig:3p2_2L}, for $\kappa=0.1$, lower than $\tau_1=0.5$ which fixes the strength of $L_{\A,1}$, and the other parameters are as in the previous experiments. The results are in agreement with what expected, and governed mainly by the interaction between the parties $\Pc_{\A}$ and $\Pc_{\M}$. In all case the  members of $\Pc_\B$  are forced to vote "no", given that $\langle \rho_0\rangle_{\psi}^{(2)}$ reaches asymptotic values that are always below $0.5$, and most often even below 0.2. The case $\gamma_1>0$ and $\gamma_2>0$, panels (a)-(b), depict a complete cooperation between the party $\Pc_{\A}$ and $\Pc_{\M}$, in agreement with the previous experiment. The non cooperative attitude for  $\Pc_{\A}$ and $\Pc_{\M}$ is instead obtained in the case $\gamma_3>0$ and $\gamma_4>0$, panels (c)-(d).

\section{Conclusions}\label{sectconclusions}

We have  proposed a dynamical approach based on the GKLS equation for the analysis of the time evolution of { a quantum parliament}, whose members are asked to approve or not a certain law. In particular, we have analysed in some details those terms in the model which produce a  {\em collaborative} and a {\em non collaborative} behaviour, to discriminate between the two.  Our approach is deduced form the idea that a small-system is influenced by the external environment, and each state of the system, representing the decision of a generic member of a party, evolves with the aforementioned GKLS equation. In particular, the various members (the small-system) can be influenced by some leaders' influence (the environment), with the possibility that the members of different parties can interact between them (in a collaborative/ non collaborative way).
We have supposed that the interactions between members are described by a Hermitian Hamiltonian, containing reversible effects, whereas the influence of the leaders is a unidirectional effect described by  Lindblad operators. With this approach we avoid the use a non-Hermitian methods, like in \cite{BaGa18,Garga21} for different macrosystems, allowing for the standard assumptions to derive the dynamics with operators in quantum mechanics. 
It is clear that the models proposed here can be adapted to other systems, of the kind discussed in the past, and a comparison between the efficiency of the various approaches is surely interesting and worth to be carried out.
These are part of our future plans.

\end{document}